\documentclass[a4paper, amsfonts, amssymb, amsmath, reprint, showkeys, nofootinbib, twoside, onecolumn, notitlepage]{revtex4-1}

\usepackage[centering,hmargin=2cm,vmargin=2cm,lmargin=1.7cm,rmargin=1.7cm]{geometry}

\def\mnras{{MNRAS}}             
\def\prd{{Phys.~Rev.~D}}        
\def\prl{{Phys.~Rev.~Lett.}}    
\usepackage{amsmath,amstext}
\usepackage[T1]{fontenc}
\usepackage{amssymb}
\usepackage{graphicx}
\usepackage{ae,aecompl}

\DeclareFontFamily{OT1}{pzc}{}
\DeclareFontShape{OT1}{pzc}{m}{it}{<-> s * [1.10] pzcmi7t}{}
\DeclareMathAlphabet{\mathpzc}{OT1}{pzc}{m}{it}

\usepackage{hyperref}
\usepackage{amsmath}
\usepackage{amssymb}
\usepackage{mathtools}
\usepackage{bm}
\usepackage{cleveref}
\usepackage{tensor}
\usepackage{braket}
\usepackage{enumitem}
\usepackage{mhchem}
\usepackage{amsthm}
\usepackage{nccmath}
\usepackage{color}

\newcommand{\spc}{\quad \quad \quad}

\def\be{\begin{equation}}
\def\ee{\end{equation}}
\def\beq{\begin{eqnarray}}
\def\eeq{\end{eqnarray}}

\begin{document}
\title{Mapping GENERIC hydrodynamics into Carter's multifluid theory}
\author{L.~Gavassino}

\affiliation{
Department of Mathematics, Vanderbilt University, Nashville, TN, USA
}

\begin{abstract}
We show that the GENERIC model for relativistic heat conduction is a multifluid of Carter. This allows one to compute the multifluid constitutive relations directly from the GENERIC formalism. As a quick application, we prove that, in the limit of infinite heat conductivity, GENERIC heat conduction reduces to the relativistic two-fluid model for superfluidity. This surprising ``crossover'' is a consequence of relativistic causality: If diffusion happens too fast, all the diffusing charge cumulates on the surface of the lightcone, and it eventually travels at the speed of light like a wave. Our analysis is non-perturbative, and it is carried out in the fully non-linear regime.
\end{abstract} 

\maketitle

\section{Introduction}\label{intro}
Most of the relativistic hydrodynamic theories currently in use rely on some form of near-equilibrium expansion. For example, the original Israel-Stewart theory arises from a perturbative expansion for small viscous fluxes \cite{Israel_Stewart_1979,Hishcock1983,Israel_2009_inbook,Salazar2020}, while BRSSS \cite{Baier2008} and BDNK \cite{BemficaDNDefinitivo2020} arise from an expansion for small gradients. DNMR arises from a subtle combination of the two expansions above \cite{Denicol2012Boltzmann}. While these approaches have the advantage of being systematic and rigorous, the resulting fluid models usually have a limited formal regime of applicability, and they may not capture all those intrinsically non-linear phenomena that real-world substances are known to exhibit. This kind of problem is well-known to material engineers, since in our everyday life we deal with many non-Newtonian fluids that easily escape a Navier-Stokes description \cite{Tropea_book}.  Indeed, most foods and toiletry products explore dynamical regimes where the shear stress component, say, $\Pi_{12}$ grows like a \textit{fractional} power of the symmetrized velocity gradient $\sigma_{12}=\partial_{(1}v_{2)}$ \cite{Steffe_book}. For example, the shear stress of ketchup at 30.5$^o$C exhibits the following non-Newtonian behavior:
\begin{equation}
    \Pi_{12} \propto |\sigma_{12}|^{0.136} \spc \text{for } \quad |\sigma_{12}| \in [0.1,100] \text{ s}^{-1} \, .
\end{equation}
Clearly, features of this kind are hardly captured by conventional gradient expansions.

Since the rheological properties of hadronic and quark matter are poorly known, we cannot exclude that non-perturbative corrections to the constitutive relations might be important for neutron star mergers and heavy-ion phenomenology. If that were the case, relativistic viscous hydrodynamics as we know it would be non-reliable for most practical applications, and we would need a full-fledged theory of relativistic rheology. 
At present, a fully general and rigorous theory of this kind exists only in the linear regime \cite{GavassinoUniversality2023}, and we are still unable to describe non-perturbative phenomena systematically. Still, some isolated ``non-linear rheological models'' have been formulated, such as Anisotropic Hydrodynamics \cite{Strickland:2014pga,Alqahtani:2017mhy}, which models shear viscosity non-perturbatively, and Hydro+ \cite{Stephanov:2017ghc,GavassinoFarbulk2023}, which models bulk viscosity non-perturbatively. The question of how to model non-perturbative corrections to heat conduction is still an open problem, and it is the subject of investigation in the present article.

First, a bit of history. The interplay between heat conduction and relativity had a long (and controversial) development. The source of all problems is a technicality: If $E{=}mc^2$, then, ``flow of energy'' = ``density of momentum'' (i.e. $T^{j0}=T^{0j}$ \cite{MTW_book}). As a consequence, heat carries inertia, and it is practically impossible to disentangle exchanges of heat from macroscopic accelerations \cite{VanKampen1968}. This is a natural consequence of the relativistic duality between energy and momentum, but failure to recognize it has generated some confusion over the definition of thermal equilibrium in relativity \cite{GavassinoTermometri}. This subtlety is also at the origin of the (unphysical) heat-driven accelerations predicted by the Eckart theory \cite{GavassinoLyapunov_2020}, and of the ambiguity between the Eckart and the Landau-Lifshitz hydrodynamic frames \cite{Salazar2020}. 

The modern theory of relativistic heat conduction was pioneered by Carter \cite{noto_rel,Lopez09}. He started from the observation that, if the heat carries inertia, then the four-momentum of the fluid must be ``redistributed'' between the entropy current $s^\alpha$ and the particle current $n^\alpha$. By symmetry, the most general way of distributing the four-momentum is
\begin{equation}\label{fuxi}
   T^{\alpha \beta}= Pg^{\alpha \beta} + \tilde{C} s^\alpha s^\beta +\tilde{A}(s^\alpha n^\beta{+}n^\alpha s^\beta)+\tilde{B} n^\alpha n^\beta \, ,
\end{equation}
where $P$, $\tilde{A}$, $\tilde{B}$, and $\tilde{C}$ are functions of $s^\alpha$ and $n^\alpha$. This gives rise to a ``two-fluid model'' for heat conduction, where dissipation is represented as friction between the two currents $s^\alpha$ and $n^\alpha$. Not surprisingly, one can show that, when the currents are almost collinear, Carter's theory reduces to the Israel-Stewart theory \cite{PriouCOMPAR1991,OlsonRegular1990,GavassinoStabilityCarter2022}. However, one can in principle extend the constitutive relations of $P$, $\tilde{A}$, $\tilde{B}$, $\tilde{C}$ to states where the angle between $s^\alpha$ and $n^\alpha$ is large, allowing one to model heat conduction in non-perturbative regimes \cite{GavassinoKhalatnikov2022}. 

Unfortunately, at present, the applicability of Carter's theory to non-perturbative regimes remains merely hypothetical. In fact, there is no concrete proposal for the constitutive relations of $P$, $\tilde{A}$, $\tilde{B}$, and $\tilde{C}$ beyond the Israel-Stewart limit. In this brief article, we present a novel and rigorous result which might be helpful in this direction. In fact, here we prove that there is an \textit{exact} change of variables that maps the GENERIC theory\footnote{GENERIC is an acronym for ``General Equation for the NonEquilibrium Reversible-Irreversible Coupling'' \cite{Grmela1997}. It is grounded on the assumption that the equations of motion of any non-equilibrium thermodynamic system, with degrees of freedom $\Psi$, can be expressed in the form $\dot{\Psi}{=}F_H(\Psi){+}F_S(\Psi)$, where $F_H$ of a reversible Hamiltonian part and $F_S$ is an irreversible (entropy-driven) Lyapunov part.} for non-linear heat conduction developed by \citet{OttingerSoloHeat1998} into Carter's multifluid theory. This implies that it is possible to derive the constitutive relations of Carter's theory directly from the GENERIC formalism \cite{Grmela1997,OttingerReview2018}. Since the latter was designed with the very purpose of modeling non-perturbative phenomena in complex fluids \cite{OttingerReview2018}, this might be our only way of deriving Carter's theory for heat conduction directly from microphysics.

Throughout the paper, we adopt the metric signature $(-,+,+,+)$, and work in natural units: $c=k_B=\hbar=1$.

\section{Relativistic dissipative hydrodynamics in a nutshell}\label{nutshell}

Historically, there have been two different ways of modeling dissipation in a hydrodynamic setting. One is the so-called ``gradient expansion'', where the degrees of freedom of the theory are the ordinary fluid fields $\{T,\mu,u^\alpha\}$, representing respectively temperature, chemical potential, and flow velocity (each field corresponds to a conservation law, respectively energy, particles, and momentum \cite{kovtun_lectures_2012,Glorioso2018}), and the stress-energy tensor is expanded in powers of gradients of such variables, e.g. $T^{\mu \nu}=T^{\mu \nu}(\partial^0)+T^{\mu \nu}(\partial^1)+T^{\mu \nu}(\partial^2)+\mathcal{O}(\partial^3)$. This approach goes back to Navier and Stokes \cite{landau6}, and Burnett \cite{Agrawal2020}, and it is often invoked as the main rationale for viscous hydrodynamics \cite{Baier2008,Romatschke2017,FlorkowskiReview2018}. The second is known as ``Extended Irreversible Thermodynamics'' \cite{Jou_Extended} (EIT), and it consists of adding new non-equilibrium variables besides the usual fluid ones, i.e. $\{T,\mu,u^\alpha,X,Y,Z,...\}$. This expands the state-space of the fluid, enabling dissipation when the new variables differ from their local equilibrium values. These extended fluid variables fully characterize \textit{all} the properties of the fluid at a given event $x^\mu$. Thus, no gradient corrections are needed, i.e. $T^{\mu \nu}=T^{\mu \nu}(T,\mu,u^\alpha,X,Y,Z,...)$. This approach was pioneered by Maxwell \cite{Roylance2001}, Cattaneo \cite{cattaneo1958}, and Muller \cite{Muller_book}, and it finds many applications in the theory of viscoelasticity \cite{Frenkel_book,landau7,FindleyBook,Steffe_book,BAGGIOLI20201}.

To better visualize the difference between the two approaches, consider the problem of deriving hydrodynamics from kinetic theory. In this setting, the goal is to express the one-particle distribution function $f_p(x^\mu)$ in terms of a set of fluid fields $\Psi(x^\mu)=\{\Psi_1(x^\mu),\Psi_2(x^\mu),...\}$, i.e. $f_p=f_p[\Psi]$. Then, the relativistic Boltzmann equation \cite{MTW_book,cercignani_book}
\begin{equation}\label{boltzuz}
    \dfrac{d}{d\lambda} f_p \bigg|_{\text{Geodesic }\big(x(\lambda),p(\lambda) \big)} = \text{``Collision integral''} \, ,
\end{equation}
written in terms of $\Psi$, can be used to derive some approximate equations of motion for the fluid fields. 

In the ideal-fluid limit, the distribution is locally Fermi-Dirac for Fermions, or Bose-Einstein for Bosons, so that
\begin{equation}\label{fermibose}
    f_p(T,\mu,u^\alpha)= \dfrac{1}{e^{-(\mu{+}u_\alpha p^\alpha)/T}{\pm} 1} \spc (+\text{ for Fermions}; \, \,-\text{ for Bosons}) \, .
\end{equation}
This clearly shows that, for ideal fluids, the natural field degrees of freedom are $\Psi=\{ T,\mu,u^\alpha \}$. Indeed, plugging \eqref{fermibose} into \eqref{boltzuz}, it is straightforward to derive the equations of motion of the ideal fluid \cite{huang_book}.

In the dissipative case, one needs to model the non-equilibrium deviations from \eqref{fermibose}. In the gradient-expansion approach, one still assumes that $\Psi=\{ T,\mu,u^\alpha \}$, but now $f_p$ becomes a \textit{functional} of $\Psi$, i.e.
\begin{equation}\label{fribuz}
    f_p[T,\mu,u^\alpha]= \dfrac{1}{e^{-(\mu{+}u_\alpha p^\alpha+\phi_p[T,\mu,u^\alpha])/T}{\pm} 1} \spc (+\text{ for Fermions}; \, \,-\text{ for Bosons}) \, ,    
\end{equation}
where $\phi_p[\Psi]$ depends on the value of $\Psi$ in a (possibly large) neighborhood $\mathcal{U}(x^\mu)$ of the event $x^\mu$ under consideration. Now, if $T$, $\mu$, and $u^\alpha$ are analytic across $\mathcal{U}(x^\mu)$, their behavior in the neighborhood is uniquely determined by the values of all their infinite derivatives at $x^\mu$ \cite{Rauch_book}, so that
\begin{equation}
    \phi_p[\Psi]=\phi_p(\Psi,\partial_\mu \Psi,\partial_\mu \partial_\nu \Psi,...) \, .
\end{equation}
Expanding $\phi_p$ in powers of derivatives leads to viscous theories such as relativistic Navier-Stokes \cite{Eckart40,landau6}, BDNK \cite{Bemfica2019_conformal1,BemficaDNDefinitivo2020,Kovtun2019}, BRSSS \cite{Baier2008}, and IRED \cite{Wagner:2022ayd}. Adding up all the terms of the series of $\phi_p$ usually leads to a divergence. However, with the aid of Borel resummation, one can reconstruct, in some cases, the exact value of $\phi_p$ \cite{Heller:2015dha,Heller:2016rtz,Romatsche2018,Spalinski2018}.

The approach of Extended Irreversible Thermodynamics is different. Here, one parameterizes the quantity $\phi_p$ in equation \eqref{fribuz} in terms of some structural fields $X_n(x^\mu)$, which determine its analytical form. For example, one may express $\phi_p(x^\mu)$ as a series $\phi_p(x^\mu)=\sum_n X_n(x^\mu)g_n(p)$, where $g_n(p)$ form a basis of $L^2(\text{``Mass hyperboloid''})$ \cite{MTW_book}. This results in a theory with hydrodynamic fields $\Psi=\{T,\mu,u^\alpha,X_n \}$, and distribution function
\begin{equation}\label{EEEEEiTT}
     f_p(T,\mu,u^\alpha,X_n)= \dfrac{1}{e^{-[\mu{+}u_\alpha p^\alpha+\phi_p(X_n)]/T}{\pm} 1} \spc (+\text{ for Fermions}; \, \,-\text{ for Bosons}) \, .   
\end{equation}
No information about the local gradients enters the formula for $f_p$ (and thus that of $T^{\mu \nu}$). Instead, the gradients appear when we plug \eqref{EEEEEiTT} into \eqref{boltzuz}. This produces some equations of motion for $X_n$, where the local gradients appear as sources. If we truncate the sequence of $X_n$ (invoking some power counting scheme), and keep only the most relevant variables, we recover transient hydrodynamic theories, such as Israel-Stewart hydrodynamics \cite{Israel_Stewart_1979}, DNMR \cite{Denicol_Relaxation_2011} (in part \cite{Denicol2012Boltzmann}), the divergence-type theory \cite{Liu1986}, and maximum entropy hydrodynamics \cite{Chattopadhyay:2023hpd}. Carter's theory and the GENERIC adopt the EIT approach, too. Therefore, we will stick here to interpretation \eqref{EEEEEiTT}. 

We remark that one should not view \eqref{fribuz} and \eqref{EEEEEiTT} as in contradiction with each other. Rather, they are different techniques for approximating the same complicated system. As such, they focus on different aspects of the problem. The mathematical relationship between the two viewpoints has been studied systematically in several works \cite{LindblomRelaxation1996,Geroch1995,HellerSingulant2022,WagnerGavassino2023}.

\section{Outline of Carter's multifluid theory}

In this section, we briefly review Carter's multifluid theory. The reader can see \cite{carter1991,carter92,Carter_starting_point,noto_rel,
andersson2007review,Termo,GavassinoRadiazione} for an overview of the foundations of the formalism. For the purposes of this paper, it will be enough to adopt the generating function approach developed in \cite{GavassinoKhalatnikov2022}. Also, we will work in the ``pressure \& momenta'' representation \cite{Prix_single_vortex,GavassinoStabilityCarter2022}.

\subsection{Multifluid constitutive relations}

Our fundamental degrees of freedom are two (unconstrained) covectors, $w_\mu$ and $\xi_\mu$, which are called \textit{momenta} of the fluid. The number of algebraic degrees of freedom is therefore $8=5 \, (\text{equilibrium})+3\, (\text{heat flux})$, as in the Israel-Stewart theory for heat conduction\footnote{Carter's theory does not pick a specific a hydrodynamic frame, i.e., it is not explicit on the definitions of $T$, $\mu$, and $u^\mu$ out of equilibrium. However, the ``natural'' frames take the form $u^\mu =a w^\mu +b \xi^\mu$ (for some state functions $a$, $b$), $T=-u^\mu w_\mu$, and $\mu=-u^\mu \xi_\mu$ \cite{Carter_starting_point,Lopez09,Termo}.}. We anticipate here that $w_\mu$ coincides with the homonymous structural variable introduced by \citet{OttingerSoloHeat1998}. The multifluid equation of state of the system is an expression of the form
\begin{equation}
P=P(w_\mu,\xi_\mu,g^{\alpha \beta}) \, ,
\end{equation}
where $P$ is the kinematic pressure of the multifluid transversal to the heat flux, and $g^{\alpha \beta}$ is the inverse metric tensor. The constitutive relations for the entropy current $s^\mu$, the conserved particle current $n^\mu$, and the (symmetric) stress-energy tensor $T^{\mu \nu}$ are computed from the equation of state through the following differential:
\begin{equation}\label{generating}
\dfrac{d(\sqrt{-g}P)}{\sqrt{-g}} = -s^\mu dw_\mu -n^\mu d\xi_\mu -\dfrac{1}{2} T_{\alpha \beta} dg^{\alpha \beta} \, .
\end{equation}
Lorentz covariance implies that $P$ can depend only on the scalars $w^\mu w_\mu$, $w^\mu \xi_\mu$, and $\xi^\mu \xi_\mu$ (recall that, within Extended Irreversible Thermodynamics, gradients do not enter the constitutive relations, see section \ref{nutshell}). Therefore, we can express the differential of $P$ in the following form \cite{Prix_single_vortex}:
\begin{equation}\label{dPippo}
-2dP =C \, d(g^{\alpha \beta}w_\alpha w_\beta) +2A \,  d(g^{\alpha \beta}\xi_\alpha w_\beta) + B \, d(g^{\alpha \beta}\xi_\alpha \xi_\beta) \, ,   
\end{equation}
where $C$, $A$, and $B$ are some non-equilibrium thermodynamic scalars. Plugging \eqref{dPippo} into \eqref{generating}, we obtain
\begin{equation}\label{pluto}
\begin{split}
\begin{bmatrix}
s^\mu \\
n^\mu \\
\end{bmatrix}
={}&  
\begin{bmatrix}
C & A \\
A & B \\
\end{bmatrix}
\begin{bmatrix}
w^\mu \\
\xi^\mu \\
\end{bmatrix} \, , \\
T^{\alpha \beta} = {}& Pg^{\alpha \beta} + s^\alpha w^\beta + n^\alpha \xi^\beta \, . \\
\end{split}
\end{equation} 
Note that the stress-energy tensor above has the form \eqref{fuxi}, and it is therefore symmetric, even if not manifestly so. The inverse of the $2\times 2$ matrix in the first equation of \eqref{pluto} is called \textit{entrainment matrix}, and it enters the quadratic form in equation \eqref{fuxi}. Both the entrainment matrix and its inverse are are positive definite, by stability \cite{GavassinoStabilityCarter2022}. This results into three notable inequalities: $C>0$, $B>0$, and $CB-A^2>0$.


\subsection{Multifluid field equations}

The fields of the hydrodynamic theory are $\varphi_i=\{w_\mu,\xi_\mu \}$, which constitute 8 algebraic degrees of freedom. Hence, we need 8 independent equations of motion. Out of these, 5 are the conservation laws, $\nabla_\mu n^\mu=0$, and $\nabla_\mu T\indices{^\mu _\nu}=0$. Invoking \eqref{dPippo} and \eqref{pluto}, the energy-momentum conservation can be equivalently rewritten in the form \cite{Carter_starting_point}
\begin{equation}\label{enomentum}
\nabla_\mu T\indices{^\mu _\nu}=2s^\mu \nabla_{[\mu} w_{\nu]}+2n^\mu \nabla_{[\mu}\xi_{\nu]}+w_\nu \nabla_\mu s^\mu=0 \, .
\end{equation}
To complete the system, we need 3 additional equations of motion. The multifluid theory does not provide a unique choice for such equation. Instead, it provides us with some reasonable options, inspired from geometry and thermodynamics \cite{noto_rel,Lopez09,GavassinoRadiazione,GavassinoKhalatnikov2022}. A rather convenient option is
\begin{equation}\label{lastequations}
2n^\mu \nabla_{[\mu}w_{\nu]} = -\Xi \, \bigg(g\indices{^\mu _\nu} - \dfrac{n^\mu n_\nu}{n^\alpha n_\alpha} \bigg) w_\mu \, ,
\end{equation}
which plays the role of an Israel-Stewart-type relaxation equation for the heat flux \cite{GavassinoKhalatnikov2022}, where $\Xi \geq 0$ is a transport coefficient.
Indeed, if we contract \eqref{enomentum} with $n^\nu$, and use equation \eqref{lastequations}, we find that the fluid produces entropy:
\begin{equation}\label{lastequations2}
 \nabla_\mu s^\mu = -\dfrac{\Xi}{n^\lambda w_\lambda} \, \bigg(g\indices{^\mu _\nu} - \dfrac{n^\mu n_\nu}{n^\alpha n_\alpha}  \bigg)s^\nu w_\mu  \, .
\end{equation}
The non-negativity of the entropy production follows from the observation that the term in the round brackets on the right-hand side is the spacelike projector orthogonal to $n^\mu$, and $Bs^\nu=(BC-A^2)w^\nu+An^\nu$, so that
\begin{equation}
   \nabla_\mu s^\mu = -\dfrac{\Xi}{n^\lambda w_\lambda} \, \dfrac{BC-A^2}{B} \, \bigg(g\indices{^\mu _\nu} - \dfrac{n^\mu n_\nu}{n^\alpha n_\alpha}  \bigg)w^\nu w_\mu \geq 0 \, .
\end{equation}

\section{Recovering the GENERIC theory}\label{RGH}

Let us now show that the multifluid model described above is mathematically equivalent to the GENERIC theory for heat conduction constructed by \citet{OttingerSoloHeat1998}, just expressed using different notation. 

\subsection{Decomposition of tensors}\label{geenr}

The GENERIC theory adopts the Eckart frame, meaning that it defines a preferred four-velocity $u^\mu$, aligned with the particle current: $n^\mu = n u^\mu$. The temperature and chemical potential are defined through the equations $T=-u^\mu w_\mu$ and $\mu = -u^\mu \xi_\mu$. The vectors $u^\mu$ and $w^\mu$ are treated as primary vectors, meaning that all other tensors are geometrically decomposed as linear combinations of tensor products of $\{u^\mu, w^\mu,g^{\alpha \beta} \}$. For example, $s^\mu$ and $\xi^\mu$ are expressed as
\begin{equation}\label{agrid}
\begin{split}
s^\mu ={}& (s{-}\sigma T)u^\mu +\sigma w^\mu \, ,\\
\xi^\mu ={}& \dfrac{\varepsilon{+}P{-}\sigma T^2}{n} u^\mu - \dfrac{s{-}\sigma T}{n} w^\mu \, . \\
\end{split}
\end{equation}
Comparing \eqref{agrid} with \eqref{pluto}, we obtain the dictionary relations below:
\begin{equation}
s=n AB^{-1}+\sigma T \spc \sigma=C-A^2 B^{-1} \spc \varepsilon=n^2 B^{-1}-P+\sigma T^2 \, .
\end{equation}
Since the inverse entrainment matrix is positive definite by stability, we have that $\sigma>0$, and $\varepsilon+P > 0$. If we plug \eqref{agrid} into the multifluid formula for the stress energy tensor, as given in \eqref{pluto}, we obtain
\begin{equation}\label{dementor}
T^{\alpha \beta}= Pg^{\alpha \beta}+(\varepsilon{+}P{-}\sigma T^2 )u^\alpha u^\beta+\sigma w^\alpha w^\beta \, .
\end{equation}
In accordance with the Eckart frame definition, we have that $n$ and $\varepsilon$ are indeed the particle and energy density as measured in the reference frame defined by $u^\mu$, namely $n=-n^\mu u_\mu$ and $\varepsilon=T^{\alpha \beta}u_\alpha u_\beta$.

\subsection{First law of thermodynamics}

In the GENERIC representation, the differential \eqref{generating} at fixed metric components becomes
\begin{equation}
dP =-\big[(s-\sigma T)u^\mu +\sigma w^\mu \big]dw_\mu -nu^\mu d\xi_\mu \, .
\end{equation}
We can use the Leibnitz rule to rewrite this differential as follows:
\begin{equation}\label{pianoforte}
dP=sdT+nd\mu -\dfrac{\sigma}{2}d(w^\mu w_\mu{+}T^2) +(sw_\mu{-}\sigma T w_\mu {+}n\xi_\mu)du^\mu \, .
\end{equation}
The second equation of \eqref{agrid} can be rewritten in the form $sw_\mu -\sigma T w_\mu +n\xi_\mu =(\varepsilon{+}P{-}\sigma T^2)u_\mu$, so that the last term in \eqref{pianoforte} vanishes. Furthermore, if we contract the second equation of \eqref{agrid} with $u_\mu$, we obtain $\varepsilon+P=T s +\mu n$, in agreement with equation (50) of \cite{OttingerSoloHeat1998}. This allows us to derive the following differential:
\begin{equation}\label{FirST}
d\varepsilon = Tds+\mu dn+ \dfrac{\sigma}{2} d(w^\mu w_\mu{+}T^2) \, .
\end{equation}
Our equations \eqref{agrid}, \eqref{dementor}, and \eqref{FirST} are indeed consistent with equations (39), (45), (54), (55), and (65) of \cite{OttingerSoloHeat1998}\footnote{Note that \citet{OttingerSoloHeat1998} splits the energy into rest mass part plus internal contributions, while we are combining them together. Thus, the notation ``$\rho_{\text{f}} c^2+\varepsilon_{\text{f}}$'' of \cite{OttingerSoloHeat1998} becomes just $\varepsilon$ in our notation.}. This shows that the constitutive relations of the GENERIC theory can be recovered from the multifluid formalism, by simply expressing all equations in the Eckart frame.

\subsection{Recovering the GENERIC field equations}

To complete our analysis, we only need to verify that equations  \eqref{lastequations} and \eqref{lastequations2} are indeed the dissipative equations of motion of the GENERIC theory. To this end, let us set $\Xi=n/\tau$, where $\tau$ is the heat relaxation time. Then, \eqref{lastequations} and \eqref{lastequations2} become
\begin{equation}\label{frugale}
\begin{split}
& u^\mu (\nabla_\mu w_\nu -\nabla_\nu w_\mu)=-\dfrac{1}{\tau} (w_\nu-Tu_\nu)\, ,\\
& T\nabla_\mu s^\mu = \dfrac{\sigma}{\tau} (w^\mu w_\mu +T^2) \, , \\
\end{split}
\end{equation}
which indeed coincide with equations (73) and (75) of \cite{OttingerSoloHeat1998}. This completes the correspondence between the GENERIC and the multifluid theory. Note that no approximation was made, meaning that the ``dictionary'' is exact in the fully non-linear regime. For completeness, we summarise the GENERIC equations in appendix \ref{appenDonae}.

\section{Infinite heat conduction becomes superfluidity}\label{ernvmcoefc}

As an application of the above result, let us show that, if $\Xi \rightarrow 0$ (which corresponds to sending the heat conductivity to infinity), the GENERIC theory becomes indistinguishable from the two-fluid model for superfluidity of  \citet{Son2001}, for certain initial conditions. Our starting point is the established fact \cite{Termo} that a superfluid is a multilfuid of Carter, where the equation of motion \eqref{lastequations} is replaced by the irrotationality condition $\nabla_{[\mu}\xi_{\nu]}=0$, see appendix \ref{RSH} for the proof. Hence, we just need to set $\Xi\equiv 0$ in \eqref{lastequations} and \eqref{lastequations2}, and verify that superfluid dynamics is recovered.

\subsection{Rigorous proof}

For $\Xi=0$, the GENERIC equations of motion reduce to
\begin{equation}\label{gabemo}
\begin{split}
&\nabla_\mu n^\mu = 0 \, , \\
&\nabla_\mu s^\mu = 0 \, , \\
&2n^\mu \nabla_{[\mu}w_{\nu]} = 0 \, , \\
& 2s^\mu \nabla_{[\mu} w_{\nu]}+2n^\mu \nabla_{[\mu}\xi_{\nu]}=0 \, . \\
\end{split}
\end{equation}
As we can see, the fluid is reversible. However, it is not exactly superfluid, because the third equation differs from the superfluid requirement that $\nabla_{[\mu}\xi_{\nu]}$ should vanish. However, let us assume that, on an initial Cauchy surface,
\begin{equation}\label{initialcondition}
\nabla_{[\mu}w_{\nu]}=\nabla_{[\mu}\xi_{\nu]}=0 \, .
\end{equation}
Then, inserting the third equation of \eqref{gabemo} into Cartan's magic formula,
\begin{equation}
\mathcal{L}_{\textbf{n}}(dw)= \iota_{\textbf{n}} d^2 w+d(\iota_\textbf{n}dw)=0 \, ,
\end{equation}
we find that $\nabla_{[\mu}w_{\nu]}$ vanishes across the whole spacetime. Hence, the fourth equation of \eqref{gabemo} becomes $2n^\mu \nabla_{[\mu}\xi_{\nu]}=0$. Again, using the initial condition \eqref{initialcondition}, Cartan's magic formula,
\begin{equation}
\mathcal{L}_{\textbf{n}}(d\xi)= \iota_{\textbf{n}} d^2 \xi+d(\iota_\textbf{n}d\xi)=0 \, ,
\end{equation}
guarantees that also $\nabla_{[\mu}\xi_{\nu]}$ vanishes everywhere. In conclusion, we have that, under the initial condition \eqref{initialcondition}, GENERIC fluids with infinite heat conductivity are \textit{exact} solutions of the superfluid equations of motion.

The above proof is a multifluid generalization of the Helmholtz theorem, according to which an inviscid fluid that is initially irrotational remains irrotational at later times.
Unfortunately, $\nabla_{[\mu}w_{\nu]}$ is not necessarily zero in a superfluid. Hence, not all superfluid configurations are reproduced by the GENERIC theory. However, in most statistical mechanical calculations, one starts from a fluid in homogeneous equilibrium, which is then disturbed by a gravitational perturbation to the metric. Under these conditions, \eqref{initialcondition} holds, and the heat conductive fluid responds like a superfluid to fully non-linear disturbances.

\subsection{Intuitive explanation}

The above result is the non-linear generalization of a recent analysis carried out in \cite{GavassinoUniversalityII2023}. There, it was found that, close to homogeneous equilibrium, the infinite conductivity limit of the linearised Israel-Stewart theory for heat conduction \cite{Israel_Stewart_1979} is mathematically equivalent to the linearised Landau two-fluid model for superfluidity  (see \cite{landau6}, \S 141). This seems rather counterintuitive, as one would expect that, by increasing the diffusivity of the system, we increase dissipation, which is at odds with the non-dissipative character of a superflow. To understand this phenomenon, let us consider a simplified one-dimensional model. 

For incompressible flows, the thermal sector of the linearised Israel-Stewart theory can be approximated by the Cattaneo equation,
\begin{equation}\label{Dv}
\dfrac{\partial^2_t T}{v^2} + \dfrac{\partial_t T}{D} = \partial^2_x T \, ,
\end{equation}
where $T$ is the temperature, $D$ is the thermal diffusivity (proportional to the heat conductivity), and $v$ is the speed of propagation of information. Now, if $D \rightarrow +\infty$, the second term in \eqref{Dv} vanishes. On the other hand, the speed $v$ cannot diverge together with $D$, because the principle of causality requires that $v^2 \in [0,1]$, and causality is necessary for stability \cite{GavassinoCausality2021,GavassinoSuperlum2021,
GavassinoBounds2023}. Therefore, equation \eqref{Dv} must become a non-dissipative wave equation:
\begin{equation}\label{w2}
\partial^2_t T = v^2 \partial^2_x T \, .
\end{equation}
This is telling us that, when $D=+\infty$, the system becomes reversible, and heat travels like an undamped wave. As it turns out, equation \eqref{w2} is also the propagation equation of the second sound in superfluids \cite{landau6}. 

\begin{figure}
\begin{center}
	\includegraphics[width=0.6\textwidth]{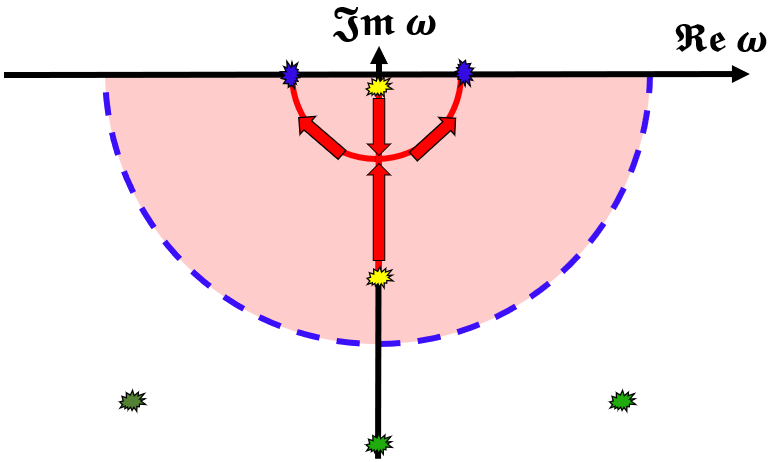}
	\caption{Hydrodynamic frequencies $\omega_{1,2}$ of the Cattaneo model \eqref{Dv} when we increase $D$ at constant $k$ and $v$. For small $D$, there is one diffusive mode (upper yellow star) and a purely relaxing non-hydrodynamic mode (lower yellow star). As we increase $D$, the modes approach each other, until they overlap. Then, both frequencies acquire a real part, and they travel on a circle of radius $vk$. When $D=+\infty$, the frequencies ``collapse'' on the real axis (blue stars), and the system becomes non-dissipative. The pink region defines the (assumed) regime of applicability of the hydrodynamic theory \cite{Denicol_Relaxation_2011,Grozdanov2019,
GavassinoQuasi2022}. Outside of it, there may be infinite other dispersion relations (green stars), which are not captured by the hydrodynamic theory.}
	\label{fig:poli}
	\end{center}
\end{figure}

Figure \ref{fig:poli} shows the behavior of the dispersion relations of the Cattaneo theory as $D$ increases, keeping $v$ constant\footnote{Sending $D {\rightarrow}+\infty$ while holding $v$ constant may seem to be a rather artificial limit, but it is actually the most natural one. In fact, the speed of information is given by the equation $v^2=D/\tau$. Both $D$ and $\tau$ usually scale like $g^{-1}$, where $g$ is the coupling constant determining the strength of the mechanism at the origin of the relaxation (e.g. the particle cross-section in a gas \cite{huang_book,Denicol2012Boltzmann,Rocha:2023hts}, or the quasi-particle interaction in a quantum liquid \cite{landau10,khalatnikov_book}). Thus, if we send $g$ to zero, $D$ diverges, while $v$ remains constant, being independent of $g$.}. The key to resolving the apparent paradox of large diffusion without dissipation is causality. If we increase the diffusivity, heat spreads faster and faster (the diffusive mode initially moves down in the complex plane). However, energy cannot spread superluminally. For this reason, at some point (when the diffusive mode meets the non-hydrodynamic mode), heat will no longer be able to diffuse in a Gaussian manner, and it will develop a front, which carries energy like a wave (both frequencies acquire a real part). If we further increase the diffusivity, we just cumulate more heat on the front. Eventually, all the energy travels like a wave, and the propagation becomes reversible.

We remark that a similar type of ``second sound'' has already been observed in solids \cite{Chester1963,SecondSound1964,Enz1974,DingChenBai2022}. Given that most liquids exhibit solid-like behaviors at short distances (see, e.g., \cite{BAGGIOLI20201,Kume2021}), it may be possible to observe the transition in figure \ref{fig:poli} experimentally also in fluids. To this end, one needs to make the relaxation time $\tau$ (and therefore $D$) parametrically large, effectively turning on wave-like dynamics. This can be achieved through confinement \cite{YuBaggioli2023}.

\section{Conclusions}

We have proved that there is an exact duality between the GENERIC theory for heat conduction \cite{OttingerSoloHeat1998} and Carter's two-fluid theory \cite{noto_rel}. This mathematical correspondence can be used to derive the constitutive relations of relativistic fluids non-perturbatively, directly within the GENERIC framework \cite{Grmela1997}. Below, we propose a formal setup that should allow one to compute the GENERIC constitutive relations from microphysics.

Let us consider a fluid in hydrostatic equilibrium, crossed by a time-independent flow of heat in the direction $x^1$. Assuming invariance under translations in the directions $x^2$ and $x^3$, we can solve the first equation of \eqref{frugale}, and we find
that $w_1=-\tau \partial_1 T$. Furthermore, from equation \eqref{dementor}, we have that the heat flux in the GENERIC theory is given by $q^1=T^{01}=T\sigma w^1$. Considering that $\sigma$ is itself a (even) function of $w^1$, we obtain the following identity:
\begin{equation}
    \dfrac{q^1}{T} =- \sigma \big(\tau \partial_1 T\big) \, \tau \partial_1 T  \, .
\end{equation}
This can be interpreted as the non-linear generalization of Fourier's law. If we have a microscopic formula for $q^1$ in this simple setup (derived, e.g., from kinetic theory), we can use it to obtain the constitutive relation $\sigma \big(\tau \partial_1 T\big)$ by direct matching. From this, we can reconstruct the GENERIC equation of state in the non-perturbative regime.

For illustrative purposes, let us consider an analytical (1+1)-dimensional kinetic toy model. 

Assume that the heat is transported by quasi-particles with speed 1 and average lifetime $\tau=\text{const}$. Then, we can write the following (approximate) kinetic formula for the heat flux in the origin:
\begin{equation}
    q^1(0)= -\int_0^{+\infty} \big[\Sigma(x)-\Sigma(-x) \big]e^{-x/\tau} dx
\end{equation}
This equation tells us that the fluid element at location $\pm x$ sends towards the origin quasiparticles that move with velocity $\mp 1$, each of which has probability $e^{-x/\tau}$ of reaching the origin. The source term $\Sigma$ needs to be prescribed from microphysics. If we choose, say, $\Sigma=e^{T(x)} \approx e^{T(0)+x\partial_1T(0)}$, the integral converges for $|\tau \partial_1 T|<1$, and we find
\begin{equation}
    T\sigma = \dfrac{2\tau e^T}{1-(\tau \partial_1 T)^2} \, .
\end{equation}
This formula is by no means realistic, but it serves as proof of principle, to show that it is possible to compute deviations from Fourier's law using non-perturbative techniques. Such deviations can then be implemented into Carter's theory using the duality with the GENERIC theory.

We remark that, while the GENERIC theory has the advantage of being more closely related to statistical mechanics, Carter's framework is a remarkably versatile formalism, due to its close connection with exterior calculus \cite{Prix_single_vortex,Carter_Prix_Magnus,Carter_defects2000,langlois98}. Thus, the present duality may also be useful for those who wish to study the mathematical properties of the GENERIC theory, as we did in section \ref{ernvmcoefc}.

\section*{Acknowledgements}

L.G. is partially supported by a Vanderbilt's Seeding Success Grant. This research was supported in part by the National Science Foundation under Grant No. PHY-1748958. I would like to thank Matteo Baggioli for useful insights on the possibility of measuring the crossover from heat conduction to superfluidity in real-world liquids.
\appendix

\section{Brief overview of the GENERIC theory}\label{appenDonae}

The natural degrees of freedom of the GENERIC theory are $\Psi=\{n,u^\mu,w_\mu \}$. The temperature is defined by the relation $T=-u^\mu w_\mu$. The equation of state, in these variables, is an equation of the form $\mathfrak{F}=\mathfrak{F}(T,n,w^\mu w_\mu)$, where $\mathfrak{F}$ is the non-equilibrium free energy density, as measured in the local rest frame. The non-equilibrium entropy density $s$ and chemical potential $\mu$ are defined by the differential
\begin{equation}
    d\mathfrak{F}=-sdT+\mu dn +\dfrac{\sigma}{2}d(w^\mu w_\mu +T^2) \, .
\end{equation}
The energy density is $\varepsilon=\mathfrak{F}+Ts$ and the pressure is $P=-\mathfrak{F}+\mu n$. These are Legendre transforms of the free energy, and their differentials are \eqref{FirST} and \eqref{pianoforte} respectively. Furthermore, they are related to each other by the Euler relation $\varepsilon+P=Ts+\mu n$. The constitutive relations for stress-energy tensor, entropy current, and particle current are, respectively,
\begin{equation}
    \begin{split}
T^{\mu \nu}={}& Pg^{\mu \nu}+(\varepsilon{+}P{-}\sigma T^2 )u^\mu u^\nu+\sigma w^\mu w^\nu \, , \\
s^\mu ={}& (s{-}\sigma T)u^\mu +\sigma w^\mu \, ,\\
n^\mu ={}& n u^\mu \, . \\
    \end{split}
\end{equation}
The equations of motion are the conservation laws $\nabla_\mu T^{\mu \nu}=\nabla_\mu n^\mu =0$, and
\begin{equation}
   u^\mu (\nabla_\mu w_\nu -\nabla_\nu w_\mu)=-\dfrac{1}{\tau} (w_\nu-Tu_\nu)\, .\\ 
\end{equation}

\section{Recovering the relativistic two-fluid model for superfluidity}\label{RSH}

The fact that the multifluid formalism can be used to describe superfluids is well known \cite{Carter_starting_point,cool1995,carter92}. Indeed, Carter's theory is currently used to model superfluid phenomena in neutron stars \cite{langlois98,sourie_glitch2017,
Geo2020,RauWasserman2020,GavassinoIordanskii2021}. Also, the correspondence with the framework of \cite{Son2001,HoloSup2009,Battacharya2011,Bhattacharya2014,
JensenKovtun2012} has already been established in \cite{Termo,andersson2007review}. However, it is instructive to have a more direct proof within the generating function approach, which is what we provide in this appendix. The analysis is somewhat specular to that carried out in section \ref{RGH}, with a different choice of preferred four-velocity $u^\mu$. Also, we must warn the reader that the quantities $\varepsilon$, $n$, $T$, and $\mu$ introduced in this appendix are not the same as those introduced in section \ref{RGH}. Indeed, one should regard this appendix and section \ref{RGH} as two \textit{alternative} ways of representing Carter multifluids, arising from different choices of $u^\mu$ (i.e. different ``hydrodynamic frames'').

\subsection{Decomposition of tensors}

The framework of \cite{Son2001,HoloSup2009,Battacharya2011,Bhattacharya2014,
JensenKovtun2012} defines a preferred four-velocity $u^\mu$, which is aligned with the entropy current: $s^\mu =su^\mu$. Such velocity is also known as the \textit{normal velocity} \cite{landau6}, and it is different from the Eckart-frame velocity introduced in section \ref{geenr}. Then, one treats $u^\mu$ and $\xi_\mu$ as the  ``primary vectors'' of the theory, meaning that all other tensors are geometrically decomposed as linear combinations of tensor products of $\{u^\mu, \xi^\mu,g^{\alpha \beta} \}$. For example, $n^\mu$ and $w^\mu$ are expressed as
\begin{equation}\label{inuovi}
\begin{split}
n^\mu ={}& nu^\mu +f^2 \xi^\mu \, ,\\
w^\mu ={}& \dfrac{\varepsilon+P}{s} u^\mu -\dfrac{n}{s} \xi^\mu \, .\\
\end{split}
\end{equation}
Comparing \eqref{inuovi} with \eqref{pluto}, we immediately obtain the following ``dictionary relations'':
\begin{equation}
n= s A C^{-1} \spc f^2=B-A^2 C^{-1} \spc \varepsilon = s^2 C^{-1}-P \, .
\end{equation}
Since the inverse entrainment matrix is positive definite (by stability \cite{GavassinoStabilityCarter2022}) we have that $n \geq 0$, $f>0$, $\varepsilon+P \geq 0$. If we plug \eqref{inuovi} into the multifluid formula for the stress energy tensor, as given in \eqref{pluto}, we obtain
\begin{equation}\label{stressone}
T^{\alpha \beta}=P g^{\alpha \beta} + (\varepsilon +P)u^\alpha u^\beta + f^2 \xi^\alpha \xi^\beta \, .
\end{equation}
It is important to keep in mind that $n$ and $\varepsilon$ are not $-n^\mu u_\mu$ and $T^{\alpha \beta}u_\alpha u_\beta$, namely the particle and entropy density as measured in the normal rest frame. Instead, $n$ and $\varepsilon$ constitute only the ``normal parts'' of the respective densities.

\subsection{Superfluid Gibbs-Duhem equation}\label{GBBS}

Let us define the temperature and the chemical potential as $T=-u^\mu w_\mu$ and $\mu =-u^\mu \xi_\mu$ \cite{Termo}. The latter definition agrees with equation (12) of \cite{HoloSup2009}. Then, contracting the second equation of \eqref{inuovi} with $u_\mu$, we obtain $\varepsilon+P=Ts+\mu n$, which is also consistent with \cite{HoloSup2009}. Now, let us focus on the differential \eqref{generating}, at fixed metric components,
\begin{equation}
dP = -s u^\mu dw_\mu -(nu^\mu +f^2 \xi^\mu) d\xi_\mu \, .
\end{equation} 
This differential can be manipulated using the Leibnitz rule, $dT{=}{-}u^\mu dw_\mu{-}w_\mu du^\mu$ and $d\mu{=}{-}u^\mu d\xi_\mu{-}\xi_\mu du^\mu$, giving
\begin{equation}
dP=sdT+nd\mu-f^2 \xi^\mu d\xi_\mu +(sw_\mu +n\xi_\mu) du^\mu \, .
\end{equation}
On the other hand, equation \eqref{inuovi} implies that $sw_\mu +n\xi_\mu=(\varepsilon+P)u_\mu$, meaning that the last term vanishes identically ($u_\mu u^\mu=-1 \, \Rightarrow u_\mu du^\mu=0$). Hence, we finally obtain the superfluid Gibbs-Duhem equation
\begin{equation}\label{Gibsus}
dP = sdT+nd\mu -\dfrac{f^2}{2} d(\xi^\mu \xi_\mu) \, .
\end{equation}
Assuming that $\xi_\mu$ is the gradient of the phase of the superfluid order parameter ($\xi_\mu =\nabla_\mu \varphi$), we see that our equations \eqref{inuovi}, \eqref{stressone}, and \eqref{Gibsus} agree with equations (9), (13), and (14) of \cite{HoloSup2009}, and we have that $\nabla_{[\mu}\xi_{\nu]}=0$. This completes our proof that the constitutive relations of Carter's multifluids are equivalent to the constitutive relations of the relativistic two-fluid model for superfluidity, as given in \cite{Son2001,HoloSup2009,Battacharya2011,Bhattacharya2014,
JensenKovtun2012}.

\bibliography{Biblio}

\label{lastpage}

\end{document}